\documentclass[a4paper,10pt,twoside]{article}
\RequirePackage{fancyhdr}
\DeclareMathSizes{10}{10}{6}{5}
\newcommand{\thanksmark}{\textsuperscript{\,\rm{*}}}

\newcommand{\ruledown}{\vspace*{-0.5cm}\hfill
\noindent{\vspace*{-0.2cm}\lower.38cm\hbox{\rule{0.2pt}{0.4cm}}\rule{8.35cm}
{0.2pt}}\vspace*{-3.5mm}}
\makeatletter
\renewcommand{\thefootnote}{\fnsymbol{footnote}}
\renewcommand{\thanks}[1]{\thanksmark
\protected@xdef\@thanks{\@thanks
\protect\footnotetext[0]{\hspace*{-6pt}$*$\;#1}}}
\newcounter{email}
\newcommand{\email}[1]{%
\protected@xdef\@thanks{\@thanks%
\protect\footnotetext[0]{\hspace*{-8pt}\arabic{email})\,{E-mail:\,}#1}}%
\stepcounter{email}}%
\renewcommand\footnoterule{%
\kern 1\p@
\hrule \@width37mm
\kern 8\p@}
\renewcommand\@makefntext[1]{%
\parindent 1em%
\noindent
\hb@xt@2em{\hss\@makefnmark}#1}
\renewcommand\maketitle{\par%
\begingroup
\renewcommand\thefootnote{\@fnsymbol\c@footnote}%
\def\@makefnmark{\rlap{\@textsuperscript{\normalfont\@thefnmark}}}%
\long\def\@makefntext##1{\parindent 1em\noindent
\hb@xt@2em{%
\hss\@textsuperscript{\normalfont\@thefnmark}}##1}%
\if@twocolumn
\ifnum \col@number=\@ne
\@maketitle
\else
\twocolumn[\@maketitle]%
\fi
\else
\newpage
\global\@topnum\z@   
\@maketitle
\fi
\@thanks
\endgroup
\setcounter{footnote}{0}%
\global\let\thanks\relax
\global\let\maketitle\relax
\global\let\@maketitle\relax
\global\let\@thanks\@empty
\global\let\@author\@empty
\global\let\@date\@empty
\global\let\@title\@empty
\global\let\title\relax
\global\let\author\relax
\global\let\date\relax
\global\let\and\relax}
\renewcommand\@maketitle{%
\begin{center}%
\let \footnote \thanks
\vspace*{0.5em}
{\LARGE\bf \@title \par}%
{\normalsize%
\lineskip .5em%
\vskip 2em%
\begin{tabular}[t]{c}%
\@author%
\end{tabular}}%
\end{center}}%
\makeatother
\newcommand{\address}[1]{%
\begin{center}%
\vskip -1em%
\begin{center}%
{\footnotesize #1}%
\end{center}%
\end{center}%
}%
\renewenvironment{abstract}%
{\small\vspace{0.5mm}%
\list{}{\rightmargin 2em%
\leftmargin 2em}%
\item{}{\bf Abstract}\hspace*{0.5em}\relax}%
{\endlist}
\newenvironment{keyword}%
{\small%
\list{}{\rightmargin 2em%
\leftmargin 2em}%
\item{}{\bf Key~words}\hspace*{0.5em}\relax }%
{\endlist%
}%
\newenvironment{pacs}%
{\small%
\list{}{\rightmargin 2em%
\leftmargin 2em}%
\item{}{\bf PACS}\hspace*{0.5em}\relax }%
{\endlist%
\vskip 6mm}%
\newcommand{\acknowledgments}[1]{%
{\vspace{10pt} \it #1}%
}
\makeatletter
\renewcommand \thesection {\bf\@arabic\c@section}
\renewcommand\section{\@startsection {section}{1}{\z@}%
{5mm \@plus.2ex \@minus .2ex}%
{5mm \@plus.2ex \@minus .2ex}%
{\normalfont\large\bfseries}}
\renewcommand\subsection{\@startsection{subsection}{2}{\z@}%
{1.5ex \@plus .2ex}%
{1.5ex \@plus .2ex}%
{\normalfont\bfseries}}
\renewcommand\subsubsection{\renewcommand \thesection {\@arabic\c@section}
\@startsection{subsubsection}{3}{\z@}%
{0.5ex}%
{0.5ex}%
{\normalfont}}
\renewcommand{\@biblabel}[1]{#1}
\renewcommand\refname{{\normalsize\bf References}}
\renewenvironment{thebibliography}[1]
{\vspace{10pt}\noindent\refname%
\@mkboth{\MakeUppercase\refname}{\MakeUppercase\refname}%
\footnotesize
\list{\@biblabel{\@arabic\c@enumiv}}%
{\settowidth\labelwidth{\@biblabel{#1}}%
\leftmargin\labelwidth
\advance\leftmargin\labelsep
\@openbib@code
\usecounter{enumiv}%
\let\p@enumiv\@empty
\renewcommand\theenumiv{\@arabic\c@enumiv}}%
\setlength{\itemsep}{0mm}
\setlength{\labelsep}{0.8em}
\setlength{\parsep}{0mm}
\setlength{\parskip}{0mm}
\setlength{\topsep}{0mm}
\setlength{\partopsep}{0mm}
\clubpenalty4000
\@clubpenalty \clubpenalty
\widowpenalty4000%
\sfcode`\.\@m}
{\def\@noitemerr
{\@latex@warning{Empty `thebibliography' environment}}%
\endlist}
\newenvironment{mylabc}
{%
\small
\let\\\@centercr
\list{}{\itemsep      \z@
\itemindent   -1em%
\listparindent0em
\leftmargin   3em
\rightmargin  2em}
\item\relax}
{\endlist}
\makeatother
\fancypagestyle{myfoot}
{%
\fancyhf{}
\fancyhead[l]{\hspace{0.5em}
\footnotesize \thepage---\pageref{LastPage}}
\fancyhead[r]{\footnotesize \today\hspace{0.5em}}
\fancyhead[c]{}
\fancyfoot[l]{}

}%
\makeatletter
\newcommand\figcaption{\def\@captype{figure}\caption}
\newcommand\tabcaption{\def\@captype{table}\caption}

\long\def\@makecaption#1#2{%
\vskip\abovecaptionskip
\sbox\@tempboxa{#1.\quad #2}%
\ifdim \wd\@tempboxa >\hsize
\begin{mylabc}
\vspace{-2mm}
{\small #1.\quad #2}%
\vskip 1mm%
\end{mylabc}\par
\else
\global \@minipagefalse
\hb@xt@\hsize{\hfil\box\@tempboxa\hfil}%
\fi
\vskip\belowcaptionskip}
\makeatother
\usepackage{multicol}
\usepackage{graphicx}
\usepackage{lastpage}
\usepackage{cite}
\usepackage{color}
\def\opschrift{Is $\Upsilon$(10580) really $\Upsilon (4S)$?}
\begin{document}
\fancyhead[co]{\footnotesize Eef van Beveren and George Rupp: \opschrift}
\title{\opschrift}
\author{%
Eef van Beveren$^{1}$\email{eef@teor.fis.uc.pt}%
and George Rupp$^{2}$\email{george@ist.utl.pt}%
}
\maketitle
\address{%
$^{1}$Centro de F\'{\i}sica Computacional,%
Departamento de F\'{\i}sica,%
Universidade de Coimbra, P-3004-516 Coimbra, Portugal\\
$^{2}$Centro de F\'{\i}sica das Interac\c{c}\~{o}es Fundamentais,%
Instituto Superior T\'{e}cnico,%
Universidade T\'{e}cnica de Lisboa, Edif\'{\i}cio Ci\^{e}ncia,%
P-1049-001 Lisboa, Portugal\\
}
\begin{abstract}
We analyse $e^-e^+$ data for $b\bar{b}$ production published
by the BABAR Collaboration, in the invariant-mass interval delimited
by the $B\bar{B}$ and $\Lambda_{b}\bar{\Lambda}_{b}$ thresholds.
In particular, we describe the $\Upsilon$(10580) enhancement, not as a
$b\bar{b}$ resonance, but rather as a threshold phenomenon due to the
opening of the $B\bar{B}$ decay channel and enhanced by the $\Upsilon(2D)$
bound-state pole not far below this threshold. The same data provide
evidence for the true $\Upsilon (4S)$ resonance, which we find at 10.735 GeV
with a width of 38 MeV. At higher energies, two more known $\Upsilon$
resonances are observed by us in the data and classified. The vital
role played in our analysis by the {\it universal confinement frequency}
\/$\omega$ is again confirmed, via a comparison with the charmonium spectrum.
\end{abstract}
\begin{multicols}{2}

\section{Introduction}
\label{introductie}

The higher bottomonium vector states,
discovered more than two decades ago,
are still today a puzzling topic of intensive research.
In Refs.~\cite{PRL54p377} and \cite{PRL54p381},
the CUSB and CLEO Collaborations, respectively, presented
the first results for the energy dependence of the
$R\left(\sigma_{\textstyle \mbox{\rm \scriptsize had}}/\sigma_{\mu\mu}\right)$
ratio above the open-bottom threshold.
Data of Ref.~\cite{PRL54p377} were observed with
the CUSB calorimetric detector operating at CESR (Cornell).
The experimental analysis resulted in evidence for structures at
$10577.4\pm 1$ MeV,
$10845\pm 20$ MeV, and
$11.02\pm 0.03$ GeV,
with total hadronic widths of
$25\pm 2.5$ MeV,
$110\pm 15$ MeV, and
$90\pm 20$ MeV, respectively.
Structures at about 10.68 and 11.2 GeV were not included
in the analysis of the CUSB Collaboration.
Data of Ref.~\cite{PRL54p381} were obtained from
the CLEO magnetic detector, also operating at CESR.
This experimental analysis resulted in evidence for structures at
$10577.5\pm 0.7\pm 4$ MeV,
$10684\pm 10\pm 8$ MeV,
$10868\pm 6\pm 5$ MeV, and
$11019\pm 5\pm 5$ MeV,
with total hadronic widths of
$20\pm 2\pm 4$ MeV,
$131\pm 27\pm 23$ MeV,
$112\pm 17\pm 23$ MeV, and
$61\pm 13\pm 22$ MeV, respectively.
A structure at about 11.2 GeV was not included
in the analysis of the CLEO Collaboration.

Here, we study data on hadron production
in electron-positron annihilation
in the invariant-mass interval between
the $B\bar{B}$ and $\Lambda_{b}\bar{\Lambda}_{b}$ thresholds,
published by the BABAR Collaboration \cite{PRL102p012001}.
In their paper, BABAR concentrates
on two of the resonances in the $b\bar{b}$ spectrum,
using data obtained by the BABAR detector at the PEP-II storage ring,
resulting in the Breit-Wigner parameters
$10876\pm 2$ MeV (mass) and $43\pm 4$ (width)
for the $\Upsilon (10860)$, and
$10996\pm 2$ MeV (mass) and $37\pm 3$ (width)
for the $\Upsilon (11020)$.
\begin{center}
\includegraphics[width=225pt]{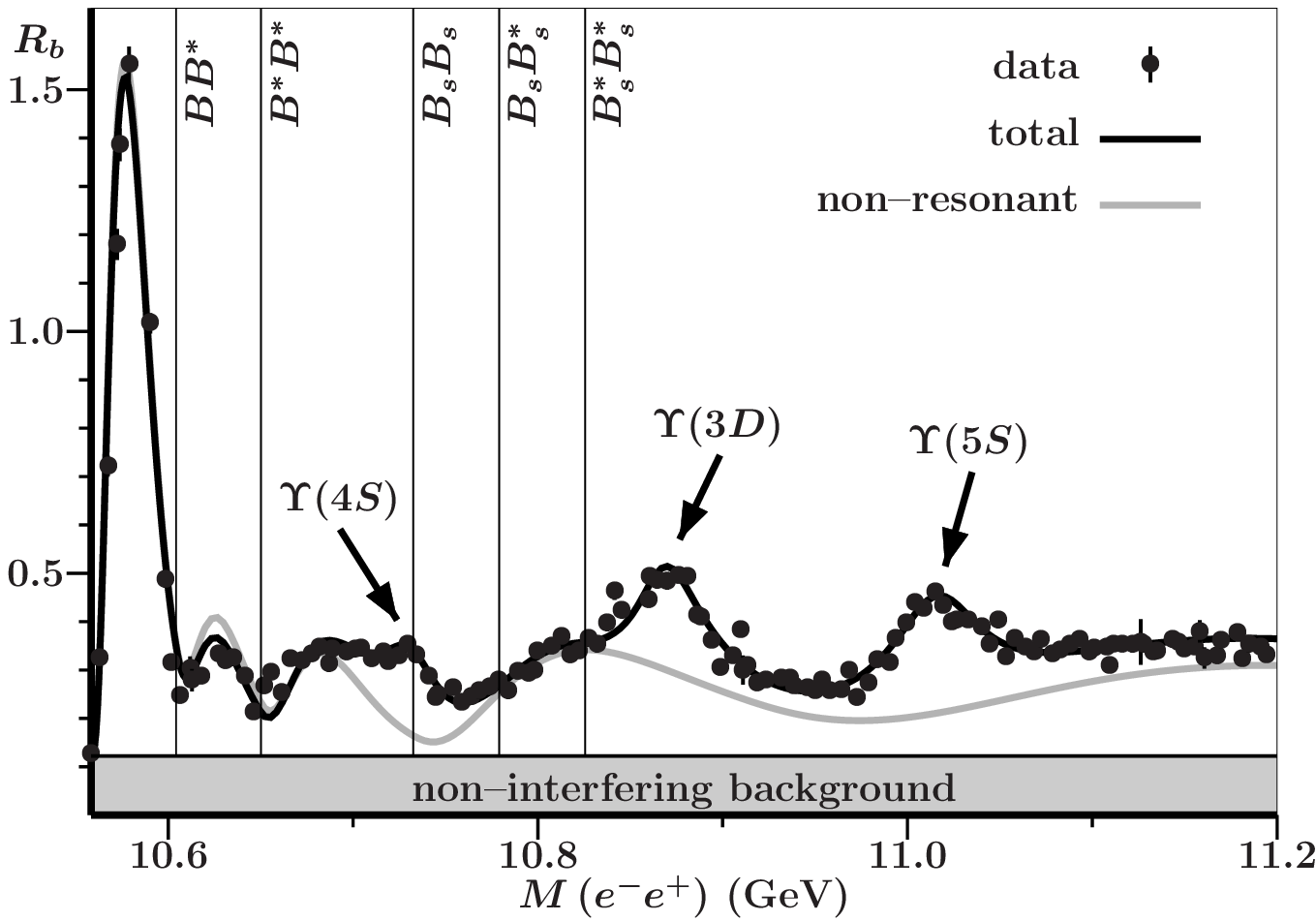}
\figcaption{\label{babarups}
BABAR data ($\bullet$ \cite{PRL102p012001}) and the results
of our analysis.}
\end{center}
The experimental line shape
of hadron production
in $e^-e^+$ annihilation
in the invariant-mass interval between
the $B\bar{B}$ and $\Lambda_{b}\bar{\Lambda}_{b}$ thresholds,
as suggested by BABAR data \cite{PRL102p012001}, and the result
of our theoretical analysis, to be discussed
in the next sections,
are shown in Fig.~\ref{babarups}.
The figure also depicts the
non-interfering background,
as well as the non-resonant contribution (solid grey curve).
Threshold positions of the $B\bar{B}^{\ast}+\bar{B}B^{\ast}$,
$B^{\ast}\bar{B}^{\ast}$, $B_{s}\bar{B}_{s}$,
$B_{s}\bar{B}_{s}^{\ast}+\bar{B}_{s}B_{s}^{\ast}$,
and $B_{s}^{\ast}\bar{B}_{s}^{\ast}$ channels are depicted
by vertical solid lines in Fig.~\ref{babarups}.
Furthermore, the central masses of the $\Upsilon (4S)$, $\Upsilon (3D)$,
and $\Upsilon (5S)$ resonances are indicated.

The BABAR results for the $\Upsilon (10860)$ and $\Upsilon (11020)$
differ substantially, in particular the widths,
from earlier results
of the CUSB \cite{PRL54p377} and CLEO \cite{PRL54p381} Collaborations,
and also from the world-average values \cite{JPG37p075021}
$10865\pm 8$ MeV (mass), $110\pm 13$ (width)
for the $\Upsilon (10860)$, and
$11019\pm 8$ MeV (mass), $79\pm 16$ (width)
for the $\Upsilon (11020)$.
Such discrepancies call for further study.

In Sect.~{\ref{ups10580}} we discuss the $\Upsilon$(10580) enhancement.
A comparison between the bound states and resonances
of the $c\bar{c}$ and $b\bar{b}$ ystems
is made in Sect.~{\ref{ccbarvsbbbar}}.
Section~{\ref{formalities}} gives details of our analysis
of the BABAR data.
Conclusions are presented in Sect.~{\ref{conclusions}}.

\section{The $\Upsilon$(10580) enhancement}
\label{ups10580}

From combined data, published by the BABAR Collaboration
in Refs.~\cite{PRD72p032005,PRL102p012001},
we observed in Ref.~\cite{PRD80p074001}
that for the enhancement just above the $B\bar{B}$ threshold
a description in terms of a wave function
with a dominant $B\bar{B}$ component is more adequate
than assuming a pole in the scattering amplitude
due to a supposed underlying $b\bar{b}$ state.
Consequently, we are convinced that it does not represent
the $\Upsilon (4S)$ $b\bar{b}$ resonance,
for the following reasons.

It is generally assumed that $B\bar{B}$ decay takes place
via the creation of a light quark pair, $u\bar{u}$ or $d\bar{d}$,
in the $b\bar{b}$ system.
However, at the creation of such a pair, there are many
possible two-meson configurations that can be formed.
Only one of these has the right quantum numbers
to develop into a $B\bar{B}$ meson pair.
But even if quantum numbers are in agreement
with the formation of a $B\bar{B}$ pair,
this does not necessarily mean that this decay will take
place.  It will only happen if a stable $B$ and a stable $\bar{B}$
can be formed. Ideally, that would be at threshold,
without any kinetic energy involved.
However, the $^{3\!}P_{0}$ mechanism
prevents decay at threshold due to a centrifugal barrier,
which, we believe, is the reason
that the signal peaks above threshold.

At higher energies, a competition mechanism sets in
between configurations that may lead to $B\bar{B}$ and
other ones, such as $B^{\ast}\bar{B}$ (or $B\bar{B}^{\ast})$.
The latter configurations may develop pairs of almost stable mesons,
when the invariant mass approaches the $B^{\ast}\bar{B}$ threshold,
which will inevitably deplete the signal from $B\bar{B}$.
One clearly sees from the BABAR data (Fig.~\ref{babarups})
that the $B\bar{B}$ signal drops to nearly zero at the $B^{\ast}\bar{B}$
threshold.  Actually, the $B^{\ast}\bar{B}$ signal itself
also drops to almost zero, namely at the $B^{\ast}\bar{B}^{\ast}$ threshold,
for the same reason.

Threshold enhancements have been described within the framework
of the Resonance-Spectrum Expansion (RSE) in Ref.~\cite{AP323p1215},
from first principles, and were further studied in Ref.~\cite{PRD80p074001}.
In the latter paper, it was shown that in electron-positron annihilation
the coupling to OZI-allowed two-meson decay channels increases from threshold,
peaks somewhat higher, and then drops again very fast.
Also, structures similar to the $\Upsilon (10580)$ have been identified
\cite{PRD80p074001}.

In this respect, an important observation was published
by the BES Collaboration in Ref.~\cite{PRL91p262001}.
To our knowledge, BES was the first to discover that the $\psi$(3770) cross
section is built up by two different amplitudes,
viz.\ a relatively broad signal and a rather narrow $c\bar{c}$ resonance.
For the narrow state, which probably corresponds to the well-established
$\psi (1D)$(3770), BES measured a central resonance position
of 3781.0$\pm$1.3$\pm$0.5 MeV
and a width of 19.3$\pm$3.1$\pm$0.1 MeV (their solution no.~2).
If the latter parameters are indeed confirmed,
it would be yet another observation
of a $q\bar{q}$ resonance width very different
from the world average
(83.9$\pm$2.4 MeV \cite{JPG37p075021} in this case),
after a similar result had been obtained
by BABAR in Ref.~\cite{PRL102p012001},
for $b\bar{b}$ resonances.
Concerning the broader charmonium structure,
the BES Collaboration indicated,
for their solution no.~2,
a central resonance position of 3762.6$\pm$11.8$\pm$0.5 MeV
and a width of 49.9$\pm$32.1$\pm$0.1 MeV.
The signal significance for the new enhancement
is 7.6$\sigma$ (solution no.~2).

In Ref.~\cite{PRD80p074001} the latter broad structure
was explained as the $D\bar{D}$ threshold enhancement.
However, in $c\bar{c}$ the situation is very different
from what one finds in $b\bar{b}$.
The $D\bar{D}$ threshold at 3.739 GeV comes out,
in the harmonic-oscillator approximation of the RSE (HORSE), just 50 MeV below
the $c\bar{c}$ confinement level at 3.789 GeV
(see Table~\ref{charmonium}), viz.\
for the degenerate $2\,{}^{3\!}S_{1}$-$1\,{}^{3\!}D_{1}$ pair.
These states get their physical masses,
which are measured in experiment,
due to the interaction generated by the meson loops.
The poles associated with the $c\bar{c}$ resonances
repel each other in such a way
that one is subject to a small mass shift,
whereas the other shifts considerably.
The higher-mass pole, mostly $1\,{}^{3\!}D_{1}$,
acquires a central resonance position that is only a few MeV
below the $c\bar{c}$ confinement level,
where it is found as the $\psi (1D)$(3770) resonance,
while the lower-mass pole, mostly $2\,{}^{3\!}S_{1}$,
comes out below the $D\bar{D}$ threshold,
as the $\psi (2S)$(3686) bound state.

In $b\bar{b}$ one has two confinement levels
that are near the $B\bar{B}$ threshold at 10.558 GeV
(see Table~\ref{upsilon}), namely the degenerate
couple $3\,{}^{3\!}S_{1}$-$2\,{}^{3\!}D_{1}$ pair 10.493 GeV, and the
degenerate couple $4\,{}^{3\!}S_{1}$--$3\,{}^{3\!}D_{1}$ at 10.873 GeV.
The former couple gives rise
to the $\Upsilon (3S)$(10355) and $\Upsilon (2D)$
bound states below the $B\bar{B}$ threshold,
due to the attraction generated by the meson loops.
Recently, in $e^{+}e^{-}\to\Upsilon (2S)\pi^{+}\pi^{-}$ data
of BABAR \cite{PRD78p112002},
possible indications were observed for
the $\Upsilon (1D)$ and $\Upsilon (2D)$ states, viz.\
at the masses 10098$\pm$5 and 10492$\pm$5 MeV
\cite{ARXIV10094097}, respectively.
Hence, the central mass of the $\Upsilon (2D)$
comes out just 1 MeV below the $b\bar{b}$ confinement level.
The latter state has its $S$-matrix pole
only about 60 MeV below the $B\bar{B}$ threshold.
Hence, it will certainly have influence
on the size of the enhancement at 10.580 GeV.

The degenerate couple $4\,{}^{3\!}S_{1}$-$3\,{}^{3\!}D_{1}$ again
produces two resonances, one of which  will have its central mass
close to the $b\bar{b}$ confinement level at 10.873 GeV.
The obvious candidate is the $\Upsilon$(10865).
The other one, viz.\ the $\Upsilon (4S)$, will be shifted towards
lower energies by the meson loops.
We will argue here that that this is not the $\Upsilon$(10580).
Actually, it would be a huge coincidence
if a resonance pole come out exactly midway between two
important threshods, viz. $B\bar{B}$ and $B^{\ast}\bar{B}$, and moreover
with an imaginary part such that the resonance peak also fits perfectly
between the two.

More than two decades ago, it seemed quite obvious that
the large enhancement just above the $B\bar{B}$ threshold
should be associated with the $\Upsilon (4S)$.
Indeed, the relativized quark model of Godfrey and Isgur
\cite{PRD32p189}, the most successful of the
typical  Coulomb-plus-linear type quarkonium models,
predicted the $\Upsilon (4S)$ state at 10.63 GeV, so
just 50 MeV too high. In view of the --- in those days ---
unpredictable threshold effects of the open-bottom decay channels,
that was a rather accurate prediction.
However, in the following we will argue
that the $\Upsilon (4S)$ $b\bar{b}$ resonance
comes out about 160 MeV higher, viz.\ at 10.735 GeV.

\section{$b\bar{b}$ spectrum in analogy with $c\bar{c}$}
\label{ccbarvsbbbar}

In the recent past, we have found possible evidence
for several higher charmonium states
\cite{EPL85p61002,PRL105p102001,ARXIV09044351,ARXIV10044368,ARXIV10053490}.
Our results are summarized in Table~\ref{charmonium}.
The masses in the first column of Table~\ref{charmonium} (HO)
are determined by
\begin{equation}
E_{q,n\ell}=2m_{q}+\omega\left( 2n+\ell +\frac{3}{2}\right)
\;\;\; ,
\label{quenched}
\end{equation}
where now $q=c$, while the charm quark mass ($m_c=1.562$ GeV)
and oscillator frequency ($\omega =0.190$ GeV)
are taken from Ref.~\cite{PRD27p1527}.
The HORSE quenched $nS$ and $(n\!-\!1)D$  $c\bar{c}$ masses are
degenerate. Unquenching the $c\bar{c}$ states
by inserting the open-charm meson-meson loops
\cite{PRD21p772,NTTP4},
also for bound states below the $D\bar{D}$ threshold,
results in a closed-form multichannel scattering amplitude,
capable of describing scattering as well as also production cross
sections, and suitable for a numerical search of its poles.
\begin{center}
\tabcaption{\label{charmonium}
Energy levels (GeV) of the HORSE quench\-ed $c\bar{c}$ spectrum (HO);
bound-state and central resonance masses (GeV)
as deduced from experiment for the $\psi$ vector states.}
\begin{tabular}{|l|ll|}
\hline
HO & $\psi (D)$ & $\psi (S)$\\
\hline
3.789 & 3.773 (1D \cite{JPG37p075021}) & 3.686 (2S \cite{JPG37p075021})\\
4.169 & 4.153 (2D \cite{JPG37p075021}) & 4.039 (3S \cite{JPG37p075021})\\
4.549 & $\approx$4.56 (3D \cite{ARXIV09044351,PRL105p102001})
& 4.421 (4S \cite{JPG37p075021}) \\
4.929 & $\approx$4.89 (4D \cite{EPL85p61002,ARXIV10053490})
& $\approx$4.81 (5S \cite{EPL85p61002,ARXIV10053490})\\
5.309 & $\approx$5.29 (5D \cite{ARXIV09044351})
& $\approx$5.13 (6S \cite{ARXIV09044351})\\
5.689 & $\approx$5.66 (6D \cite{ARXIV10044368})
&  $\approx$5.44  (7S \cite{ARXIV10044368})\\
6.069 &  -- (7D) & $\approx$5.91 (8S \cite{ARXIV10044368})\\
\hline
\end{tabular}
\end{center}
We find then that the bare $c\bar{c}$ states turn into
bound states below the $D\bar{D}$ threshold, or resonances
thereabove. The $S$ states (third column of Table~\ref{charmonium})
have central masses some 100 to 200 MeV below the unquenched levels,
whereas the $D$ states (second column of Table~\ref{charmonium})
undergo mass shifts of only a few MeV.
These mass shifts largely depend on the precise positions
of the open-charm threshold.
Results for $q=b$ ($m_{b}=4.724$ GeV \cite{PRD27p1527}),
in Eq.~(\ref{quenched}) are given in Table~\ref{upsilon}.
\begin{center}
\tabcaption{\label{upsilon}
Energy levels (GeV) of the HORSE quench\-ed $b\bar{b}$ spectrum;
bound-state and central resonance masses (GeV)
as deduced from experiment for the $\Upsilon$ vector states.}
\begin{tabular}{|l|ll|}
\hline
quenched & $\Upsilon (D)$ & $\Upsilon (S)$\\
\hline
10.113 & 10.098 (1D \cite{ARXIV10094097}) & 10.023 (2S \cite{JPG37p075021})\\
10.493 & 10.492 (2D \cite{ARXIV10094097}) & 10.355 (3S \cite{JPG37p075021})\\
10.873 & 10.865 (3D \cite{JPG37p075021}) & 10.735 (4S \cite{ARXIV09100967}) \\
11.253 &  -- (4D) & 11.019 (5S \cite{JPG37p075021})\\
\hline
\end{tabular}
\end{center}
We observe a $b\bar{b}$ spectrum which is very similar
to the $c\bar{c}$ spectrum of Table~\ref{charmonium},
just shifted towards higher masses by about 6.3 GeV.
However, our particle assignments are somewhat
different from what one finds in most of the literature.

The experimental identification of the resonance
at 10.845 GeV (CUSB) or 10.868 GeV (CLEO),
and the resonance
at 11.02 GeV (CUSB) or 11.019 GeV (CLEO),
with the $\Upsilon (5S)$ and $\Upsilon (6S)$, respectively,
was apparently inspired by the predictions of Godfrey and Isgur
\cite{PRD32p189}
for those states, at 10.88 GeV and 11.10 GeV, respectively.
However, we rather identify these resonances rather with
the $\Upsilon (3D)$ and $\Upsilon (5S)$ states, respectively,
on the basis of the level schemes in
Tables~\ref{charmonium} and \ref{upsilon}
\cite{PRD21p772,PRD27p1527}.

\section{Our analysis of the BABAR data}
\label{formalities}

The BABAR data in Ref.~\cite{PRL102p012001} concern
the $R_{b}$ ratio for all $e^{-}e^{+}$ annihilation processes
containing $b$ quarks.  Our description of the BABAR data
(see Fig.~\ref{babarups}) consists of three parts:
\begin{enumerate}
\item
A non-interfering background.
\item
Treshold enhancements interfering with the resonances.
\item
The $\Upsilon (4S)$, $\Upsilon (3D)$ and $\Upsilon (5S)$ resonances.
\end{enumerate}
In Fig.~\ref{elephit} we show details of Fig.~\ref{babarups}
for clearity.

The non-interfering background accounts for
those reactions that do not contain open-bottom pairs.
For its value we take a similar amount as suggested
by BABAR in their analysis of the heavier two resonances
\cite{PRL102p012001}.

Nonresonant threshold enhancements,
indicated by the solid grey curve in Fig.~~\ref{babarups},
are determined by several different factors;
in the first place, by the amount
of available competing configurations.
Hence, the $B^{\ast}\bar{B}$ threshold enhancement
is much less pronounced than the one for $B\bar{B}$.
Another factor is the average distance of the $b\bar{b}$ pair
at which pair production takes place.
A smaller average distance implies that the maximum
of the enhancement occurs for higher relative momenta
of the open-bottom decay products.
This phenomenon one may observe for the enhancement
at the $B_{s}\bar{B}_{s}$ threshold,
because $s\bar{s}$ pair production takes place
at smaller interquark distances than
$u\bar{u}$ and $d\bar{d}$ pair production.
In this case, the maximum is never reached, since before that
$B_{s}^{\ast}\bar{B}_{s}$ production takes over,
and similarly so  at the $B_{s}^{\ast}\bar{B}_{s}^{\ast}$ threshold.
At even higher invariant masses, several other open-bottom decays
become energetically allowed, which then results in a slowly rising
nonresonant contribution
(also see the solid grey curve in Fig.~\ref{elephit}c).

The three resonances
$\Upsilon (4S)$, $\Upsilon (3D)$, and $\Upsilon (5S)$
are parametrized by Breit-Wigner (BW) amplitudes,
which, because of relation ({\ref{quenched}}),
are linear in mass, and not quadratic
as in the standard relativistic expressions.
As a consequence, we found in Ref.~\cite{ARXIV09100967}
small deviations for the resonance pole positions, as compared
to the findings of the BABAR Collaboration.

The resonances interfere with the nonresonating
threshold-enhancement contributions,
but not with the non-interfering background.
We showed in Ref.~\cite{ARXIV09100967} that the phases
of the interferences between the resonant signals
and the nonresonating threshold-enhancement contributions
can be fully determined in the HORSE, without any freedom.
\begin{center}
\begin{tabular}{|l|}
\hline
{\bf a}\hspace{5pt}\includegraphics[width=180pt]{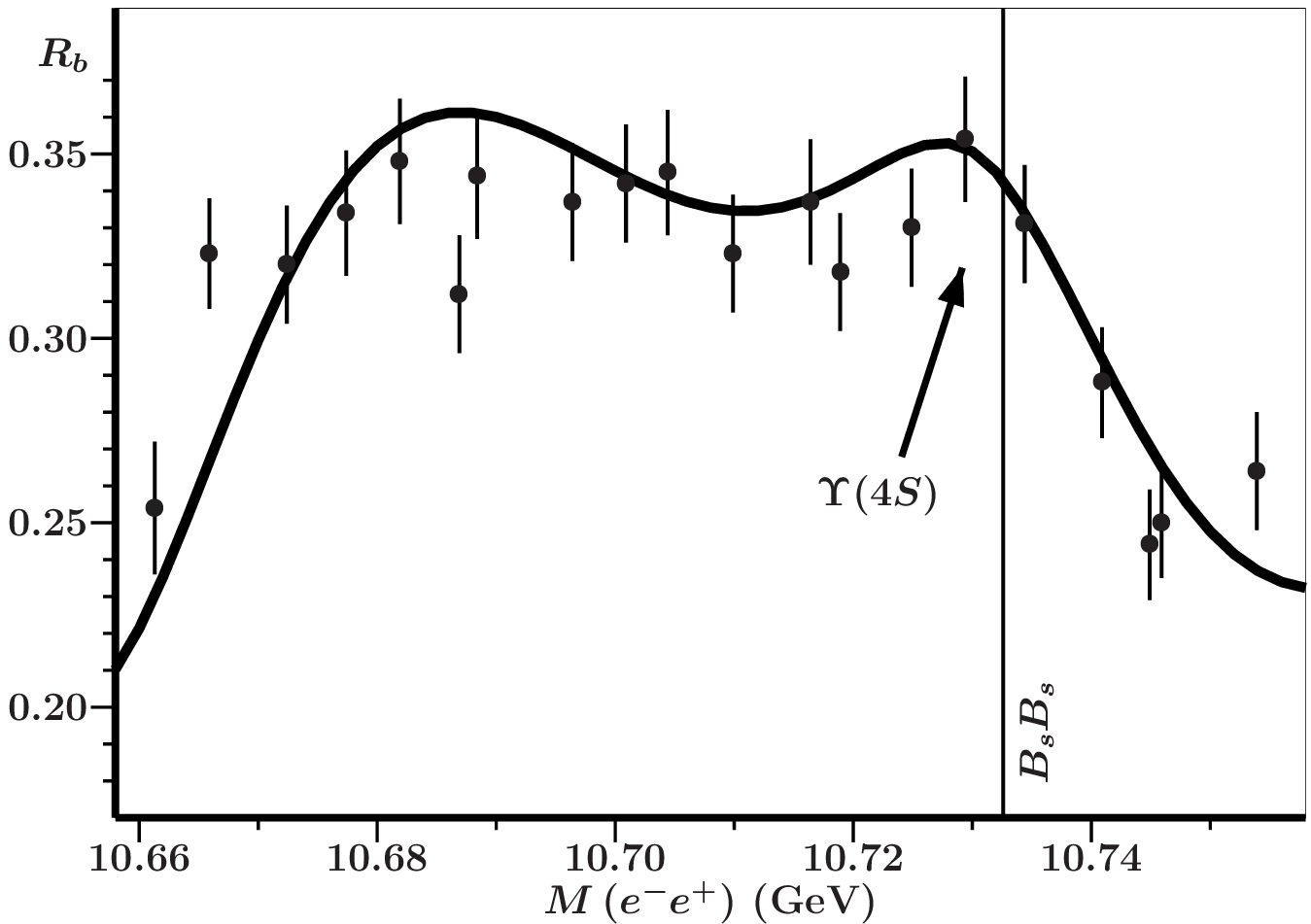}\\
\hline
{\bf b}\hspace{5pt}\includegraphics[width=180pt]{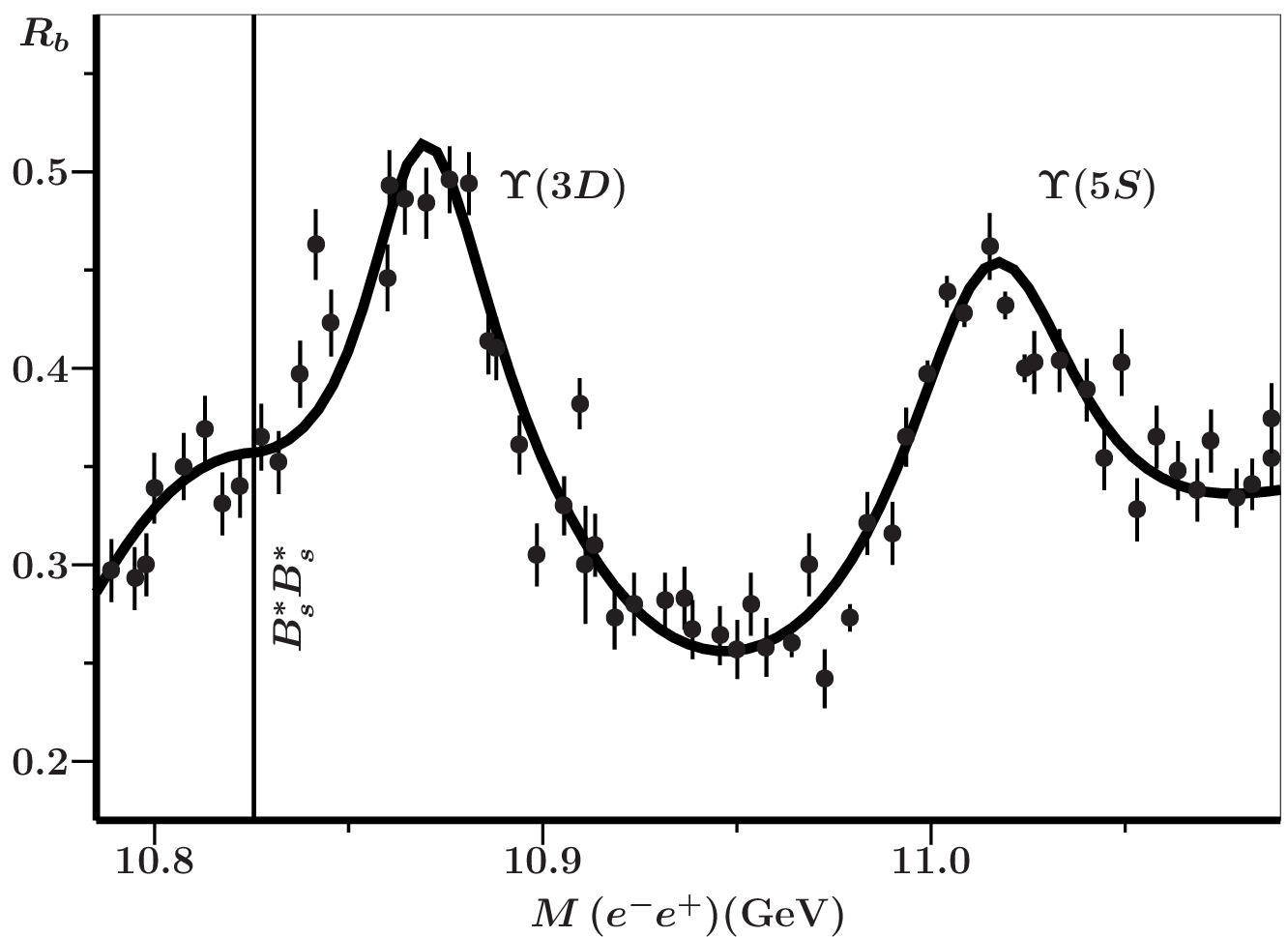}\\
\hline
{\bf c}\hspace{5pt}\includegraphics[width=180pt]{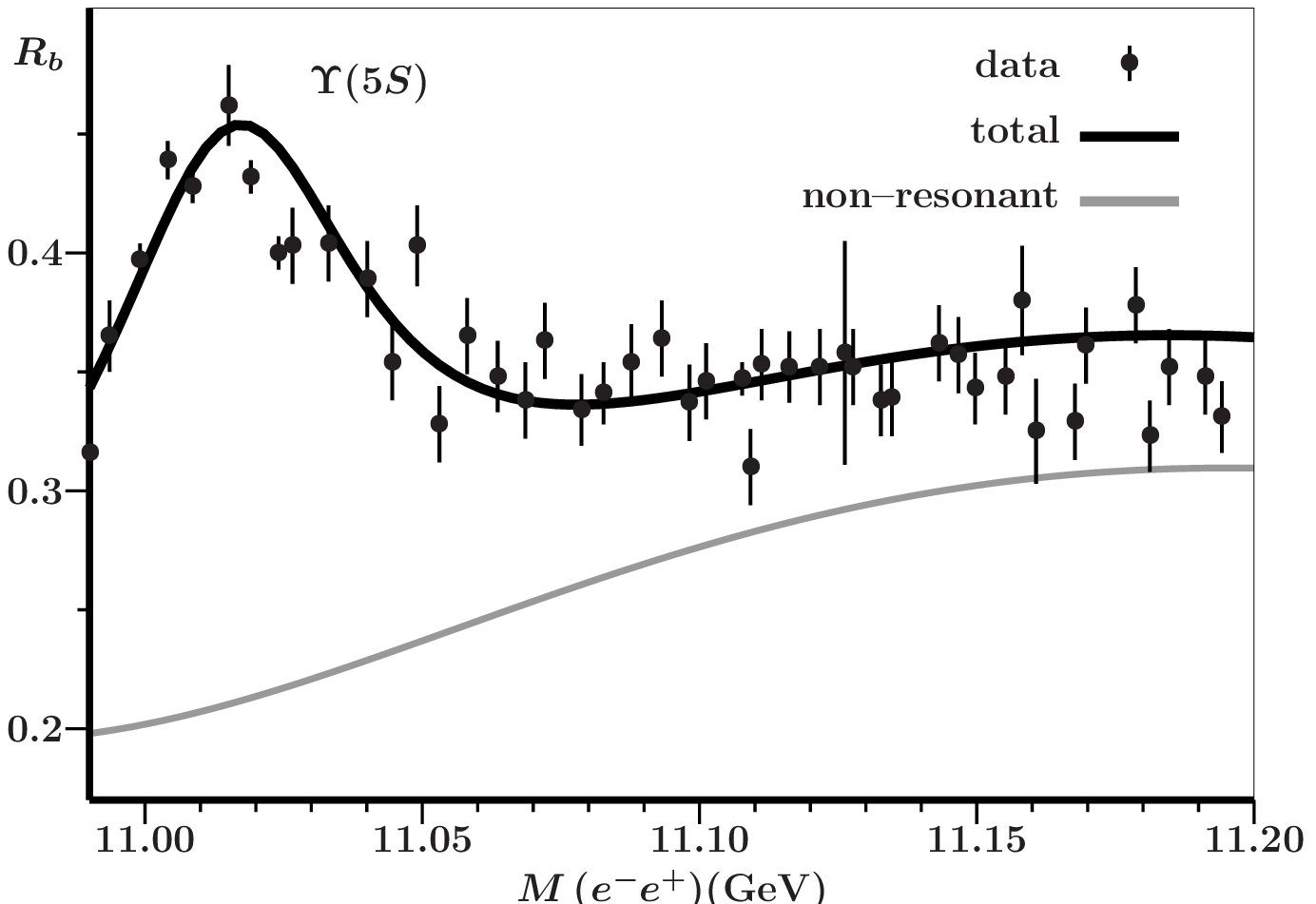}\\
\hline
\end{tabular}
\vspace{10pt}
\figcaption{\label{elephit}
Details of our results:
(a) in the $\Upsilon (4S)$ region;
(b) for the $\Upsilon (3D)$ and $\Upsilon (5S)$ resonances;
(c) the ``plateau'' (c).
Data ($\bullet$) for hadron production in electron-positron
annihilation are by BABAR \cite{PRL102p012001}.}
\end{center}

In Ref.~\cite{ARXIV09100967}, we found
real and imaginary parts for the resonance pole positions
of the $\Upsilon$(10860) and the $\Upsilon$(11020)
in reasonable agreement with those obtained by BABAR
(see Fig.~\ref{elephit}b).
However, we also found a resonant structure
at 10.735 GeV, with a width of 38 MeV,
which was not obtained in the BABAR analysis
(see Fig.~\ref{elephit}a).
We associate the latter resonance with the $\Upsilon$(4S),
as it also fits much more nicely in the level scheme of
Table~\ref{upsilon}.

In Sect.~{\ref{introductie}},
we mentioned a resonance at $10684\pm 10\pm 8$ MeV,
with a total hadronic width of $131\pm 27\pm 23$ MeV,
observed by the CLEO Collaboration \cite{PRL54p381},
which was classified as a presumable $b\bar{b}g$ hybrid.
Figure~\ref{elephit}a clearly shows that
also the BABAR data display an enhancement at that invariant mass.
However, in our analysis its origin is the nonresonant threshold
enhancement due to the $B^{\ast}\bar{B}^{\ast}$ channel, and not
the presence of a $b\bar{b}$ resonance pole,
as we will discuss in the following.

In Ref.~\cite{PRL102p012001}, the BABAR Collaboration
observed two plateaux in $R_{b}$.
The first one appears just below the $\Upsilon (4S)$, and is
depicted in Fig.~\ref{elephit}a.
As shown through our theoretical curve,
we do not associate the data with a plateau,
but rather with the ``back and shoulders'' of an ``elephant''.
Also from Fig.~\ref{babarups}
we seem to learn that neither the nonresonant contribution
nor the resonance have a particularly flat behavior
in the mass region delimited by
the $B^{\ast}\bar{B}^{\ast}$ and $B_{s}\bar{B}_{s}$ thresholds.
As for the nonresonant part,
this mass interval constitutes a window
for $B^{\ast}\bar{B}^{\ast}$ production,
which signal in part {\it carries} \/the $\Upsilon (4S)$ resonance.
Furthermore, the tail of the resonance interferes with
the nonresonant contribution, leading to the shallow dip
in between the elephant's back and shoulders.

However, the BABAR Collaboration also points at a second
plateau, above the $\Upsilon (5S)$, which we have depicted
in Fig.~\ref{elephit}c.
Here, we indeed observe a flat pattern for $R_{b}$,
which we assume to be the result of a slowly rising
nonresonant contribution (solid grey curve)
and the tail of the $\Upsilon (5S)$ resonance.

\section{Conclusions}
\label{conclusions}

Inspired by the level schemes of
Tables~\ref{charmonium} and \ref{upsilon},
we have argued that the $\Upsilon$(4S)
should be associated with the resonance at 10.735 GeV,
rather than with the the large peak
just above the $B\bar{B}$ threshold.
The latter structure is, in our analysis,
better described in terms of a wave function
with a dominant $B\bar{B}$ component,
enhanced by the nearby $\Upsilon$(2D) bound-state pole
below the $B\bar{B}$ threshold.
The vital role of the {\it universal confinement frequency}
\/$\omega =190$ MeV in analysing hadronic data
is once again supported by the results shown in Tables~\ref{charmonium}
and \ref{upsilon}.

\acknowledgments{
We are grateful for the precise measurements
and data analyses of the BABAR Collaboration
that made the present analysis possible.
This work was supported in part by
the \emph{Funda\c{c}\~{a}o para a Ci\^{e}ncia e a Tecnologia}
\/of the \emph{Minist\'{e}rio da Ci\^{e}ncia, Tecnologia e Ensino Superior}
\/of Portugal, under contract CERN/\-FP/\-83502/\-2008.}

\newcommand{\pubprt}[4]{{#1 {\bf #2}, #3 (#4)}}
\newcommand{\ertbid}[4]{[Erratum-ibid.~{#1 {\bf #2}, #3 (#4)}]}
\def\AP{Ann.\ Phys.}
\def\EPL{Europhys.\ Lett.}
\def\JPG{J.\ Phys.\ G}
\def\PRD{Phys.\ Rev.\ D}
\def\PRL{Phys.\ Rev.\ Lett.}

\end{multicols}

\begin{thebibliography}{20}
\bibitem{PRL54p377}
D.~M.~J.~Lovelock {\it et al.} [CUSB Collaboration],
\pubprt{\PRL}{54}{377}{1985}.

\bibitem{PRL54p381}
D.~Besson {\it et al.} [CLEO Collaboration],
\pubprt{\PRL}{54}{381}{1985}.

\bibitem{PRL102p012001}
B.~Aubert  [BaBar Collaboration],
\pubprt{\PRL}{102}{012001}{2009}.

\bibitem{JPG37p075021}
K. Nakamura {\it et al.} \/[Particle Data Group Collaboration],
\pubprt{\JPG}{37}{075021}{2010}.

\bibitem{PRD72p032005}
B.~Aubert  [BaBar Collaboration],
\pubprt{\PRD}{72}{032005}{2005}.

\bibitem{PRD80p074001}
E.~van Beveren and G.~Rupp,
\pubprt{\PRD}{80}{074001}{2009}.

\bibitem{AP323p1215}
E.~van Beveren and G.~Rupp,
\pubprt{\AP}{323}{1215}{2008}.

\bibitem{PRL91p262001}
S.~K.~Choi {\it et al.}  [Belle Collaboration],
\pubprt{\PRL}{91}{262001}{2003}.

\bibitem{PRD78p112002}
B.~Aubert {\it et al.}  [BABAR Collaboration],
\pubprt{\PRD}{78}{112002}{2008}.

\bibitem{ARXIV10094097}
E.~van Beveren and G.~Rupp,
arXiv:1009.4097.

\bibitem{PRD32p189}
S.~Godfrey and N.~Isgur,
\pubprt{\PRD}{32}{189}{1985}.

\bibitem{EPL85p61002}
E.~van~Beveren, X.~Liu, R.~Coimbra, and G.~Rupp,
\pubprt{\EPL}{85}{61002}{2009}.

\bibitem{PRL105p102001}
E.~van Beveren, G.~Rupp and J.~Segovia,
\pubprt{\PRL}{105}{102001}{2010}.

\bibitem{ARXIV09044351}
E.~van Beveren and G.~Rupp,
arXiv:0904.4351.

\bibitem{ARXIV10044368}
E.~van Beveren and G.~Rupp,
arXiv:1004.4368.

\bibitem{ARXIV10053490}
E.~van Beveren and G.~Rupp,
arXiv:1005.3490.

\bibitem{PRD27p1527}
E.~van Beveren, G.~Rupp, T.~A.~Rij\-ken, and C.~Dullemond,
\pubprt{\PRD}{27}{1527}{1983}.

\bibitem{PRD21p772}
E.~van Beveren, C.~Dullemond, and G.~Rupp,
\pubprt{\PRD}{21}{772}{1980}
\ertbid{\ D}{22}{787}{1980}.

\bibitem{NTTP4}
E.~van Beveren and G.~Rupp,
chapter 4 in {\it New Topics in Theoretical Physics},
Horizons in World Physics, Vol. 258, pp 47-74 (2007),
Edited by H.~F.~Arnoldus, T.~F.~George,
{\it Nova Science Publishers},
ISBN 1600213553, 9781600213557.

\bibitem{ARXIV09100967}
E.~van Beveren and G.~Rupp,
arXiv:0910.0967.
\end{thebibliography}
\end{document}